\documentclass{article}
\usepackage{graphicx}
\usepackage[centertags]{amsmath}
\usepackage{amsfonts}
\usepackage{amssymb}
\usepackage{amsthm}

\title{Quasiclassical and Quantum Systems of Angular Momentum. Part I. Group Algebras as a Framework for Quantum-Mechanical Models with Symmetries}

\author{J. J. S\l awianowski, V. Kovalchuk, A. Martens,\\ 
B. Go\l ubowska, and E. E. Ro\.zko\\
Institute of Fundamental Technological Research,\\
Polish Academy of Sciences,\\
$5^{\rm B}$, Pawi\'{n}skiego str., 02-106 Warsaw, Poland\\
e-mails: jslawian@ippt.gov.pl, vkoval@ippt.gov.pl,\\ 
amartens@ippt.gov.pl, bgolub@ippt.gov.pl, erozko@ippt.gov.pl}

\begin{document}

\maketitle
\begin{abstract}
We use the mathematical structure of group algebras and $H^{+}$-algebras for describing certain problems concerning the quantum dynamics of systems of angular momenta, including also the spin systems. The underlying groups are ${\rm SU}(2)$ and its quotient ${\rm SO}(3,\mathbb{R})$. The scheme developed is applied in two different contexts. Firstly, the purely group-algebraic framework is applied to the system of angular momenta of arbitrary origin, e.g., orbital and spin angular momenta of electrons and nucleons, systems of quantized angular momenta of rotating extended objects like molecules. The other promising area of applications is Schr\"{o}dinger quantum mechanics of rigid body with its often rather unexpected and very interesting features. Even within this Schr\"{o}dinger framework the algebras of operators related to group algebras are a very useful tool. We investigate some problems of composed systems and the quasiclassical limit obtained as the asymptotics of "large" quantum numbers, i.e., "quickly oscillating" wave functions on groups. They are related in an interesting way to geometry of the coadjoint orbits of ${\rm SU}(2)$.
\end{abstract}

Many physical systems have geometric background based on some groups or their byproducts like homogeneous spaces, Lie algebras and co-algebras, co-adjoint orbits, etc. Those group structures are relevant both for classical and quantum theories. They are basic tools for fundamental theoretical studies; they provide us also with the very effective tool for practical calculations. According to some views \cite{8}, such a purely group-theoretical background is characteristic for almost all physical models, or at least for realistic and viable ones. Let us mention a funny fact known to everybody from the process of learning or teaching quantum mechanics. After the primary struggle with elementary introduction to quantum theory, first of all to atomic and molecular physics, students are often convinced that the properties of quantum angular momentum, e.g., its composition rules, so important in atomic spectroscopy and nuclear phenomena, are some mysterious and obscure dynamical laws. And only later on, they are very surprised that there is nothing but group theory there, namely the theory of unitary irreducible representations of the three-dimensional rotation group ${\rm SO}(3,\mathbb{R})$ or its covering group ${\rm SU}(2)$ \cite{4,6,7}. And the really dynamical model assumptions are placed elsewhere. Below we concentrate on certain quantum and quasiclassical problems based on some group-theoretic apriori, first of all on the theory of quantum angular momenta and their systems, including systems of spins.

There were various views and various answers to the question: "What is quantum mechanics?" What is to be used as its proper and most adequate mathematical language? Hilbert space, rigged Hilbert space, operator algebra, wave mechanics, matrix mechanics, quantum logics, orthomodular lattices, etc.? \cite{5,1,8} We suppose there is no answer to this question, in any case, there is non as yet. Below we follow some working hypothesis, idea by F. Schroeck \cite{8} that every really fundamental and viable model in quantum, but also in classical, mechanics is always based on some apriori chosen group and its representations, cf. also \cite{21,22,20,17,18,19,6,15}. In flat-space theories, i.e., ones without gravitation, they are Euclidean, pseudo-Euclidean and affine groups (and other Lie groups, e.g., in gauge theories). When working in a manifold, i.e., when gravitation is taken into account, everything is based on the infinite-dimensional group of all diffeomorphisms. Incidentally, this group is also fundamental in certain geometric models of nonlinear quantum mechanics \cite{16}. In our treatment the main mathematical tool is theory of group algebras. And we follow the idea of W. M. Tulczyjew \cite{9} and H. Weyl \cite{6} about group algebra as the interesting, in a sense, aprioric model of quantum mechanics.

We begin with a short review of necessary mathematical preliminaries and prerequisites. This review is less than being far from completeness and it cannot be anything more here; it is just quoted to remind some elementary concepts and to fix notations. For more details and systematic theory cf. \cite{12,2,11,13,14}.

Let us begin with the concept of $H^{+}$-algebra as introduced by Ambrose \cite{12}. This is a special case of the Banach algebra with involution, but not necessarily with identity. Let us mention, incidentally, that any Banach algebra $B$ without identity may be reinterpreted as a maximal ideal in the unital Banach algebra $B\times \mathbb{C}$ with the product rule $\left(x,\lambda\right)\left(y,\mu\right):=\left(xy+\lambda y+\mu x,\lambda\mu\right)$ and the norm $\|\left(x,\lambda\right)\|:=\|x\|_{B}+|\lambda|$. Then, obviously, the element $(0,1)$ becomes the identity and $x\mapsto (x,0)$ is just the mentioned injection of $B$ into $B\times\mathbb{R}$; its image $(B,0)$ being a maximal ideal. Let us remind that the involution, denoted by $x\mapsto x^{+}$, is assumed to satisfy
\begin{equation}\label{eq_1.1}
x^{++}=x,\qquad \left(\lambda x+\mu y\right)^{+}=\overline{\lambda}x^{+}+\overline{\mu}y^{+},\qquad \left(xy\right)^{+}=y^{+}x^{+}.
\end{equation}
Often, but not necessarily, one assumes also
\begin{equation}\label{eq_1.1a}
\|xx^{+}\|=\|x\|^{2}.
\end{equation}

The algebra of bounded operators in a Hilbert space, with the usual definition of the operator norm and with the Hermitian conjugation as an involution, is a typical, very important example.

An $H^{+}$-algebra is a consistent hybrid of two structures: a Banach algebra with involution and a Hilbert space. The underlying linear space will be denoted by $B$ and the scalar product of elements $x,y\in B$ will be denoted by $\left(x,y\right)$; it is assumed to obey the usual Hilbert space axioms. Let us remind what is meant by the compatibility of those structures:
\begin{itemize}
\item The Banach and Hilbert norms are identical:
\begin{equation}\label{eq_1.2}
\|x\|^{2}=\left(x,x\right).
\end{equation}

\item The involution, referred to as Hermitian conjugation, is compatible with the Hermitian conjugation of linear operators acting in $B$. This means that for any $w\in B$ the Hermitian conjugation of the left regular translation $L_{w}:B\rightarrow B$ is identical with the left regular translation $L_{w^{+}}:B\rightarrow B$ by the involution of $w$, $\left(L_{w}\right)^{+}=L_{w^{+}}$, i.e.,
\begin{equation}\label{eq_1.3}
\left(wx,y\right)=\left(x,w^{+}y\right)
\end{equation}
for any $x,y\in B$.

\item The involution is a norm-preserving operation, i.e.,
\begin{equation}\label{eq_1.4}
\|x^{+}\|=\|x\|,
\end{equation}
for any $x\in B$
\begin{equation}\label{eq_1.5}
\left(x\neq 0\right)\quad \Rightarrow\quad x^{+}x\neq 0.
\end{equation}
\end{itemize}
All these axioms imply in particular that involution is an antiunitary operator
\begin{equation}\label{eq_1.6}
\left(x^{+},y^{+}\right)=\left(y,x\right)=\overline{\left(x,y\right)},
\end{equation}
therefore, it is an isometry of $B$ as a metric space
\begin{equation}\label{eq_1.7}
d\left(x^{+},y^{+}\right)=\|y^{+}-x^{+}\|=\|y-x\|=d\left(x,y\right);
\end{equation}
obviously, being antilinear, involution cannot be unitary. Another important consequence is that the analogue of (\ref{eq_1.3}) holds also for the right translations $R_{w}$,
\begin{equation}\label{eq_1.8}
\left(xw,y\right)=\left(x,yw^{+}\right)
\end{equation}
for any $x,y\in B$.

It is just (\ref{eq_1.3}) and (\ref{eq_1.8}) that enables one to use the same symbol for the involution in $B$ and Hermitian conjugation in ${\rm L}(B)$, the algebra of linear operators on $B$; there is no danger of confusion.

One deals very often with some special situations, when the Hilbert structure of a Banach algebra with involution is a byproduct of something more elementary, i.e., a linear functional $T:B\rightarrow \mathbb{C}$ such that
\begin{eqnarray}
T(xy)&=&T(yx),\label{eq_1.9}\\
T\left(x^{+}\right)&=&\overline{T(x)},\label{eq_1.10}\\
\left(x\neq 0\right) &\Rightarrow& \left(T\left(x^{+}x\right)>0\right).\label{eq_1.11} 
\end{eqnarray}
The scalar product is then defined as
\begin{equation}\label{eq_1.12}
\left(x,y\right)=T\left(x^{+}y\right).
\end{equation}
The most elementary example, which at the same time provides some, so-to-speak, comparison pattern for all more general situations, is the associative algebra ${\rm L}(H)$ of all linear operators acting on a finite-dimensional unitary space $H$. Scalar product of vectors $\varphi,\psi\in H$ will be denoted by $\langle\varphi|\psi\rangle$; obviously, the involution in $B={\rm L}(H)$ is defined by the usual formula
\begin{equation}\label{eq_1.13}
\langle x\varphi|\psi\rangle=\langle\varphi|x^{+}\psi\rangle
\end{equation}
for the Hermitian conjugation. Then $T$ is just the trace operation, $T(x)={\rm Tr}\ x$, and the scalar product (\ref{eq_1.12}) is given by the standard formula
\begin{equation}\label{eq_1.14}
(x,y)={\rm Tr}\left(x^{+}y\right).
\end{equation}

Let $e_{i}$ be some basic elements of $H$ and $e^{i}$ be the corresponding dual elements of the conjugate space $H^{\ast}$, thus,
\begin{equation}\label{eq_1.15}
\langle e^{i},e_{j}\rangle=e^{i}\left(e_{j}\right)=\delta^{i}{}_{j};
\end{equation}
the symbol $\langle f,u\rangle$, used as a popular abbreviation for $f(u)$, denotes the evaluation of the linear function $f\in H^{\ast}$ on the vector $u\in H$. The Gramm matrix assigned to the basis $\left(\ldots,e_{i},\ldots\right)$ has the elements
\begin{equation}\label{eq_1.16}
\Gamma_{ij}=\langle e_{i}|e_{j}\rangle,
\end{equation}
then for any vectors $u=u^{i}e_{i}$, $v=v^{j}e_{j}$ we have
\begin{equation}\label{eq_1.17}
\langle u|v\rangle=\Gamma(u,v)=\Gamma_{ij}\overline{u^{i}}v^{j}.
\end{equation}
The inverse matrix element will be denoted by $\Gamma^{ij}$,
\begin{equation}\label{eq_1.18}
\Gamma^{ik}\Gamma_{kj}=\delta^{i}{}_{j}.
\end{equation}
The scalar product of linear functions $f=f_{i}e^{i}$, $g=g_{j}e^{j}\in H^{\ast}$ is given by
\begin{equation}\label{eq_1.19}
\langle f,g\rangle=\Gamma^{ij}f_{i}\overline{g_{j}}.
\end{equation}
Obviously, this expression is correctly defined, i.e., independent on the choice of basis in $H$. Usually one prefers the choice of orthonormal bases, when
\begin{equation}\label{eq_1.20}
\Gamma_{ij}=\langle e_{i}|e_{j}\rangle=\Gamma\left(e_{i},e_{j}\right)=\delta_{ij},\qquad \Gamma^{ij}=\delta^{ij}.
\end{equation}
Any basis $\left(\ldots,e_{i},\ldots\right)$ in a linear space $H$ gives rise to the adapted basis $\left(\ldots,e_{j}{}^{i},\ldots\right)$ in ${\rm L}(H)$, where
\begin{equation}\label{eq_1.21}
e_{j}{}^{i}:=e_{j}\otimes e^{i},\qquad {\rm i.e.},\qquad e_{j}{}^{i}e_{k}=\delta^{i}{}_{k}e_{j}. 
\end{equation}
Therefore, the matrix elements of $e_{j}{}^{i}$ with respect to the basis $\left(\ldots,e_{i},\ldots\right)$ are given by
\begin{equation}\label{eq_1.22}
\left(e_{j}{}^{i}\right)^{a}{}_{b}=\delta^{i}{}_{b}\delta^{a}{}_{j}. 
\end{equation}

It is easy to see that
\begin{equation}\label{eq_1.23}
e_{j}{}^{i}e_{r}{}^{s}=\delta^{i}{}_{r}e_{j}{}^{s},\qquad {\rm Tr}\ e_{j}{}^{i}=\delta_{j}{}^{i}. 
\end{equation}
Introducing the modified basic elements
\begin{equation}\label{eq_1.24}
e_{ji}:=\Gamma_{ik}e_{j}{}^{k}=e_{j}{}^{k}\overline{\Gamma_{ki}}, 
\end{equation}
we have
\begin{eqnarray}
e_{ik}e_{lj}&=&\Gamma_{kl}e_{ij},\label{eq_1.25}\\
{\rm Tr}\left(e_{ij}\right)&=&\Gamma_{ji}=\overline{\Gamma_{ij}},\label{eq_1.26}\\
e_{ij}^{+}&=&e_{ji},\label{eq_1.27}\\
e_{i}{}^{j+}&=&\Gamma_{ia}\Gamma^{bj}e_{b}{}_{a}=
\overline{\Gamma^{jb}}e_{b}{}^{a}\overline{\Gamma_{ai}},\label{eq_1.28} 
\end{eqnarray}
where, obviously, the contravariant upper-case $\Gamma$ is reciprocal to the covariant lower-case one,
\begin{equation}\label{eq_1.29}
\Gamma^{ac}\Gamma_{cb}=\delta^{a}{}_{b},\qquad  \Gamma_{ac}\Gamma^{cb}=\delta_{a}{}^{b};
\end{equation}
it is clear that also the following holds:
\begin{equation}\label{eq_1.30}
\overline{\Gamma^{ac}}\ \overline{\Gamma_{cb}}=\delta^{a}{}_{b},\qquad  \overline{\Gamma_{ac}}\ \overline{\Gamma^{cb}}=\delta_{a}{}^{b}.
\end{equation}
The basic scalar products of operators have the following form:
\begin{eqnarray}
\left(e_{i}{}^{j},e_{a}{}^{b}\right)&=&\Gamma_{ia}\Gamma^{bj}=
\Gamma_{ia}\overline{\Gamma^{jb}},\label{eq_1.31}\\
\left(e_{ij},e_{ab}\right)&=&\Gamma_{ia}\Gamma_{bj}=
\Gamma_{ia}\overline{\Gamma_{jb}};\label{eq_1.32} 
\end{eqnarray}
these are "orthogonality" relations for the operators $e_{a}{}^{b}$, $e_{ab}$.

Some of the above formulas become remarkably simpler when the basis $\left(\ldots,e_{i},\ldots\right)$ is orthonormal,
\begin{equation}\label{eq_1.33}
\Gamma_{ij}=\delta_{ij},\qquad \Gamma^{ij}=\delta^{ij}.
\end{equation}
However, it is sometimes convenient to separate "metrical" concepts from the weaker "affine" ones, as far as possible.

It is instructive and convenient for the analysis of quantum problems to mention and make use of the Dirac notation in $H$, $B={\rm L}(H)$. The basic vectors $e_{i}$ are then denoted by $|i\rangle$ and the basic operators $e_{ij}$ are then given by
\begin{equation}\label{eq_1.34}
e_{ij}=|i\rangle\langle j|;
\end{equation}
obviously, the above notation is adapted just to the situation when the basis $\left(\ldots,e_{i},\ldots\right)$ is chosen as orthonormal. Perhaps the notation
\begin{equation}\label{eq_1.35}
P_{ij}:=|i\rangle\langle j|
\end{equation}
is then more adequate than $e_{ij}$; the diagonal elements
\begin{equation}\label{eq_1.36}
P_{i}:=P_{ii}=|i\rangle\langle i|
\end{equation}
are then orthogonal projections onto $\mathbb{C}$-one-dimensional subspaces with bases $e_{i}=|i\rangle$ and, obviously,
\begin{equation}\label{eq_1.37}
\sum_{i}|i\rangle\langle i|={\rm Id}_{H}.
\end{equation}
For a general basis we have the following completeness relation:
\begin{equation}\label{eq_1.38}
\sum_{a,b}\Gamma^{ab}e_{ab}=\sum_{a,b}\Gamma^{ab}P_{ab}={\rm Id}_{H}
\end{equation}
or using the "summation convention"
\begin{equation}\label{eq_1.39}
\Gamma^{ab}e_{ab}=\overline{\Gamma^{ba}}e_{ab}={\rm Id}_{H}.
\end{equation}

Obviously, the corresponding non-metrical "affine" completeness relation is given by
\begin{equation}\label{eq_1.40}
\sum_{a}e^{a}{}_{a}=e^{a}{}_{a}={\rm Id}_{H}.
\end{equation}

Let us notice that the operators $e_{i}{}^{i}$ (no summation convention!) are idempotents, cf. (\ref{eq_1.23}). When the basis $\left(\ldots,e_{i},\ldots\right)$ is not orthonormal, they are not Hermitian, however, they are so if the basis is orthonormal, $\Gamma_{ij}=\delta_{ij}$, cf. (\ref{eq_1.28}). Unlike this, the "diagonal" elements $e_{ii}$ (no summation convention!) are always Hermitian and in the case of orthonormal basis in $H$ they are also idempotents, cf. (\ref{eq_1.25}), (\ref{eq_1.27}), so we have then that
\begin{equation}\label{eq_1.41}
e_{ii}^{+}=e_{ii},\qquad e_{ii}e_{ii}=e_{ii}
\end{equation}
(no summation convention!). So in this case we obtain the orthonormal decomposition of the identity operator,
\begin{equation}\label{eq_1.42}
{\rm Id}_{H}=\sum_{a}e_{aa}=e^{a}{}_{a}
\end{equation}
(obviously, the summation convention meant on the extreme right-hand side).

It is easy to see that for any fixed $j$, the linear span of elements $e_{ij}$, i.e., equivalently, the linear span of elements $e_{i}{}^{j}=e_{ik}\Gamma^{jk}$, forms a minimal left ideal ${\rm L}(H)e_{ij}={\rm L}(H)e_{i}{}^{j}$ in ${\rm L}(H)$, so ${\rm L}(H)$ is a direct sum of $n=\dim H$ such ideals. One can easily show that such left ideals are generated by the operators $e_{\underline{j}}{}^{\underline{j}}$, or equivalently, by $e_{\underline{j}\underline{j}}$; no summation convention here. Let us denote those left ideals by
\begin{equation}\label{eq_1.43}
M_{j}={\rm L}(H)e_{\underline{j}\underline{j}}=
{\rm L}(H)e_{\underline{j}}{}^{\underline{j}}.
\end{equation}

Let us repeat: $M_{j}$ is the set of linear combinations of the form $\alpha^{i}e_{i}{}^{j}=\beta^{i}e_{ij}$; $\alpha$, $\beta$ are arbitrary.

Similarly, we have the minimal right ideals
\begin{equation}\label{eq_1.44}
{}_{j}M:=e_{\underline{j}\underline{j}}{\rm L}(H)=
e_{\underline{j}}{}^{\underline{j}}{\rm L}(H).
\end{equation}
They are obtained as the sets of linear combinations of the form $\alpha_{i}e_{j}{}^{i}=\beta^{i}e_{ji}$, where $\alpha$, $\beta$ again are arbitrary.

As mentioned, ${\rm L}(H)$ splits into the direct sum of ideals $M_{j}$ or ${}_{j}M$,
\begin{equation}\label{eq_1.45}
{\rm L}(H)=M_{1}\oplus\cdots\oplus M_{n}={}_{1}M\oplus\cdots\oplus {}_{n}M.
\end{equation}
This is the orthogonal splitting in the sense of (\ref{eq_1.14}); $M_{i}$ is orthogonal to $M_{j}$ if $i\neq j$, and the same is true for ${}_{i}M$, ${}_{j}M$.

Any finite-dimensional $H^{+}$-algebra $B$ is an $H^{+}$-subalgebra of the above ${\rm L}(H)$ with the induced structures. It may be uniquely decomposed into the direct sum of minimal two-sided ideals $M(\alpha)$, $\alpha=1,\ldots,k$, every one of them being isomorphic to some ${\rm L}(V)$ with the structure of $H^{+}$-algebra as described above. Therefore, $\dim M(\alpha)=n^{2}_{\alpha}$ and 
\begin{equation}\label{eq_1.46}
\sum^{k}_{\alpha=1}n^{2}_{\alpha}=\dim B.
\end{equation}
Every $M(\alpha)$ is generated by some Hermitian idempotent $e(\alpha)$,
\begin{equation}\label{eq_1.47}
M(\alpha)=Be(\alpha)B,
\end{equation}
and the following holds:
\begin{eqnarray}
e(\alpha)e(\beta)=0\quad {\rm if}\quad \alpha\neq\beta,&\quad& \left(e(\alpha),e(\beta)\right)=0\quad {\rm if}\quad \alpha\neq\beta,\\ 
e(\alpha)e(\alpha)=e(\alpha),&\quad& e(\alpha)^{+}=e(\alpha).  
\end{eqnarray}
The minimal two-sided ideals $M(\alpha)$, $M(\beta)$ are orthogonal when $\alpha\neq\beta$.

In ${\rm L}(H)$ there are only two ideals $M(\alpha)$, the improper ones, namely, ${\rm L}(H)$ itself and $\{0\}$. And, obviously, in ${\rm L}(H)$ the corresponding Hermitian idempotent is just the identity element $e={\rm Id}_{H}=e^{a}{}_{a}=\Gamma^{ab}e_{ab}$. In a general finite-dimensional $H^{+}$-algebra $B$, the minimal two-sided ideals $M(\alpha)$, being isomorphic with ${\rm L}\left(n_{\alpha},\mathbb{C}\right)\simeq\mathbb{C}^{n^{2}_{\alpha}}$, are direct sums of $n(\alpha)$ left minimal ideals $M(\alpha)_{j}$, every one of dimension $n(\alpha)$. Of course, they are also representable as direct sums of $n(\alpha)$ right minimal ideals ${}_{j}M(\alpha)$, every of dimension $n(\alpha)$. The label $j$ runs the range of naturals from 1 to $n(\alpha)$. And, on analogy to ${\rm L}(H)$ we choose some special bases $e(\alpha)_{i}{}^{j}$, $e(\alpha)_{ij}$ in $B$, where, for a fixed $\alpha$, $i$, $j$ run over the natural range from 1 to $n(\alpha)$ (to avoid the crowd of symbols we simply write $i$, $j$ instead of $i(\alpha)$, $j(\alpha)$ as we in principle should have done). Those bases are assumed to have, for a fixed $\alpha$, the properties analogous to (\ref{eq_1.23})--(\ref{eq_1.32}). More precisely, it is so when $B$ is not an abstract $H^{+}$-algebra but some $H^{+}$-subalgebra of ${\rm L}(H)$. Otherwise some comments would be necessary concerning the coefficients $\Gamma(\alpha)_{ij}$; in any case, to avoid discussion, one can put them to be the Kronecker symbols.

Nevertheless, in a general $H^{+}$-algebra there are situations when (\ref{eq_1.31}), (\ref{eq_1.32}) are modified, e.g., that for
any $\alpha$ there exists $\Gamma(\alpha)$ its own; for instance, analytically the coefficients $\Gamma(\alpha)_{rs}$
are there proportional to $\delta_{rs}$ with coefficients depending on $\alpha$. This is not the case, in (\ref{eq_1.25}), (\ref{eq_1.26}).
One can show (cf. \cite{12}) that in general the canonical $\varepsilon$-basis may be chosen in such a way that:
\begin{eqnarray}
\varepsilon(\alpha)_{ik}\varepsilon(\alpha)_{jl}&=&
\delta_{kj}\varepsilon(\alpha)_{il},\label{eq_1.9a1a}\\
\left(\varepsilon(\alpha)_{ik},\varepsilon(\alpha)_{jl}\right)&=&0 \quad {\rm unless} \quad i=j \quad {\rm and} \quad k=l,\label{eq_1.9a1b}\\
\left(\varepsilon(\alpha)_{ik},\varepsilon(\alpha)_{ik}\right)&=&
\left(\varepsilon(\alpha)_{11},\varepsilon(\alpha)_{11}\right),\label{eq_1.9a1c}\\
\varepsilon(\alpha)_{ij}^{+}&=&\varepsilon(\alpha)_{ji},\label{eq_1.9a1d}\\
\varepsilon(\alpha)&=&\sum_{i}\varepsilon(\alpha)_{ii}.\label{eq_1.9a1e}
\end{eqnarray}

The diagonal elements $\varepsilon(\alpha)_{ii}$ are irreducible Hermitian idempotents (any of them is not a sum
of two idempotents), and their sum equals the idempotent $\varepsilon(\alpha)$ generating the two-sided ideal
$M(\alpha)$. Obviously, for different $\alpha$, $\beta$ the corresponding $\varepsilon$-elements are mutually 
orthogonal and annihilate each other under multiplication,
\begin{eqnarray}
\left(\varepsilon(\alpha)_{ij},\varepsilon(\beta)_{rs}\right)&=&0 \quad {\rm if} \quad \alpha\neq \beta,\label{eq_1.9a2a}\\
\varepsilon(\alpha)_{ik}\varepsilon(\beta)_{rs}&=&0 \quad {\rm if} \quad \alpha\neq \beta,\label{eq_1.9a2b}\\
\left(\varepsilon(\alpha),\varepsilon(\alpha)\right)&=&
\left(\varepsilon(\alpha)_{11},\varepsilon(\alpha)_{11}\right)\dim M(\alpha).
\end{eqnarray}

Unlike the minimal two-sided ideals $M(\alpha)$, the left and right ideals $M(\alpha)_{i}$, ${}_{i}M(\alpha)$ are not unique.

Similarly, the basic elements $e(\alpha)_{i}{}^{j}$ or $e(\alpha)_{ij}$ are not unique. However, their "index-traces",
\begin{equation}\label{eq_1.48}
e(\alpha)=e(\alpha)^{i}{}_{i}=\Gamma(\alpha)^{ij}e(\alpha)_{ij},
\end{equation}
are unique and just coincide, as denoted, with the generating idempotents $e(\alpha)$. The "diagonal" idempotents
$e(\alpha)_{\underline{i}}{}^{\underline{i}}$ or $e(\alpha)_{\underline{i}\underline{i}}$ (no summation convention!) are not unique, however, the "trace" (\ref{eq_1.48})
is so, and their sum is just the identity element of $B$,
\begin{equation}\label{eq_1.49}
\sum_{\alpha}e(\alpha)=e.
\end{equation}
The idempotent $e(\alpha)$ is referred to as the induced unit of $M(\alpha)$.

Those roughly referred examples provide some "reference frame" for understanding the general theory.
Nevertheless, the general case, when the infinite dimension of $B$ is admitted, is much more complicated,
and many finite-dimensional analogies are misleading. Many important $H^{+}$-algebras are non-unital.
Instead of direct sums, some direct integrals of Hilbert spaces must be used. In the infinite dimension many 
structures taken from finite-dimensional operator algebras diffuse, one must oscillate between various subsets
like those of trace-class operator, Hilbert-Schmidt operators, etc.

The simplest infinite-dimensional situations are group algebras on locally compact topological groups.
In particular, group algebras on compact subgroups are relatively similar to the finite-dimensional
case, e.g., all minimal ideals are finite-dimensional.

Let $G$ be a locally compact topological group. Obviously, we are interested mainly in finite-dimensional Lie
groups, nevertheless, there is a hierarchy of structures based on more general ideas and only later on, 
on the level of applications, assuming more and more specialized concepts. Let $\mu_{l}$, $\mu_{r}$ denote
respectively the left- and right-invariant Haar measures on $G$. They are unique up to constant normalization
factors, but in general they do not coincide. Nevertheless, the right-shifted left-invariant measure
is still left-invariant, so roughly-speaking,
\begin{equation}\label{eq_1.50}
d\mu_{l}(gh)=\Delta(h)d\mu_{l}(g),
\end{equation}
where $\Delta(h)$ is a positive factor, and iterating those right transforms one can easily show that
\begin{equation}\label{eq_1.51}
\Delta(hk)=\Delta(h)\Delta(k)=\Delta(kh),
\end{equation}
so $\Delta$ is a homomorphism of $G$ into $\mathbb{R}^{+}$ as a multiplicative group. So $\mu_{l}$
and $\mu_{r}$ may be different only when $G$ does possess a nontrivial homomorphism into 
$\mathbb{R}^{+}$. If they are identical, we say that $G$ is unimodular. Compact and Abelian Lie
groups, and so their direct and semidirect products are unimodular. If $G$ is unimodular, the measure elements
$d\mu_{l}(g)=d\mu_{r}(g)$ are denoted simply by $dg$. If $G$ is unimodular then not only
\begin{equation}\label{eq_1.52}
\mu(Ah)=\mu(hA)=\mu(A),
\end{equation}
but also $\mu\left(A^{-1}\right)=\mu(A)$, for any measurable subset $A\subset G$. From now on it will be always assumed that
$G$ is unimodular, so we write symbolically
\begin{equation}\label{eq_1.53}
d(gh)=d(hg),\qquad d\left(g^{-1}\right)=dg.
\end{equation}

In the space $L^{1}(G)$ of integrable functions one defines the convolution operation
\begin{equation}\label{eq_1.54}
\left(A*B\right)(g)=\int A(h)B\left(h^{-1}g\right)dh=\int A\left(gk^{-1}\right)B(k)dk.
\end{equation}
It is defined on the total $L^{1}(G)\times L^{1}(G)$ and produces from elements of $L^{1}(G)$ the elements
of $L^{1}(G)$, so $L^{1}(G)$ is an algebra under the convolution,
\begin{equation}\label{eq_1.55}
L^{1}(G)*L^{1}(G)\subset L^{1}(G).
\end{equation}

Obviously, the convolution $*$ turns $L^{1}(G)$ into $L^{1}(G)$, so that the axioms of associative Banach 
$C^{*}$-algebra hold in $L^{1}(G)$, e.g.,
\begin{equation}\label{eq_1.56}
||f*g|| \leq ||f|| ||g||, 
\end{equation}
the $L^{1}(G)$-norm is meant, and the involution is defined as
\begin{equation}\label{eq_1.57}
\left(f^{+}\right)(x)=\overline{f\left(x^{-1}\right)}.
\end{equation}
The linear functional ${\rm Tr}$ is defined as the value at the group identity $1$:
\begin{equation}\label{eq_1.58}
{\rm Tr}(f):=f(1),
\end{equation}
the scalar product of functions on $G$ is defined as 
\begin{equation}\label{eq_1.59}
(\varphi,\psi)={\rm Tr}\left(\varphi^{+}* \psi\right)=\int \overline{\varphi}(x)\psi(x)dx.
\end{equation}
It is positive,
\begin{equation}\label{eq_1.60}
{\rm Tr}\left(\varphi^{+}\varphi\right)=\int  \overline{\varphi}(x)\varphi(x)dx>0 \quad {\rm if} \quad 0\neq \varphi \in L^{2}(G).
\end{equation}
These expressions lead us to the space $L^{2}(G)$ and algebraic structures in $L^{2}(G)$. All these structures fit
together so as to result in the structure of $H^{+}$-algebra in $L^{2}(G)$. Such structures in $L^{1}(G)$, $L^{2}(G)$
are referred to as group algebras. Everything is particularly simple when $G$ is a discrete group. Obviously, then
\begin{equation}\label{eq_1.61}
\langle \varphi, \psi \rangle=\sum_{x\in G} \overline{\varphi}(x)\psi(x).
\end{equation}
If $G$ is compact, then we usually normalize the measure such that
\begin{equation}\label{eq_1.62}
\mu(G)=1.
\end{equation}
In particular, if it is a finite, $N$-element group, we put
\begin{equation}\label{eq_1.63}
\int f(x)dx=\frac{1}{N}\sum_{x\in G}f(x).
\end{equation}
However, this normalization is not always used and not always is convenient.

If $G$ is compact, then $L^{2}(G)$ splits uniquely into the direct sum of minimal two-sided ideals $M(\alpha)$
where $\alpha$ runs over some discrete set of labels $\Omega$,
\begin{equation}\label{eq_1.64}
L^{2}(G)=\oplus_{\alpha\in \Omega}M(\alpha).
\end{equation}
These ideals are mutually orthogonal,
\begin{equation}\label{eq_1.65}
M(\alpha)\perp M(\beta) \quad {\rm if} \quad \alpha\neq \beta; \quad (F,G)=0 \quad {\rm if} \quad F\in
M(\alpha),\ G\in M(\beta),
\end{equation}
and, obviously,
\begin{equation}\label{eq_1.66}
M(\alpha)* M(\beta)=\{0\} \quad {\rm if} \quad \alpha\neq \beta; \quad F*G=0 \quad {\rm if} \quad F\in
M(\alpha),\ G\in M(\beta).
\end{equation}
They are generated by Hermitian idempotents $\varepsilon(\alpha)$, therefore,
\begin{equation}\label{eq_1.67}
M(\alpha)=\varepsilon(\alpha)*L^{2}(G)=L^{2}(G)*\varepsilon(\alpha),
\end{equation}
and, obviously, the following holds:
\begin{equation}\label{eq_1.68}
\left(\varepsilon(\alpha), \varepsilon(\beta)\right)=0 \quad {\rm if} \quad  \alpha\neq \beta, \quad 
\varepsilon(\alpha)* \varepsilon(\beta)=
\delta_{\alpha\beta}\varepsilon(\beta)n^{2}(\beta)
\end{equation}
(no summation convention in the last expression).

The convolution with $\varepsilon(\alpha)$ acts as the orthogonal projection of $L^{2}(G)$ onto $M(\alpha)$;
in particular, it is a generated unit of $M(\alpha)$:
\begin{equation}\label{eq_1.69}
\varepsilon(\alpha)*F=F*\varepsilon(\alpha)=F
\end{equation}
for any $F\in M(\alpha)$. If $G$ is a finite group with $N$ elements, then $L^{1}(G)=L^{2}(G)$ is a unital algebra under convolution.
Its identity $\varepsilon$ is proportional to the "delta" function
\begin{equation}\label{eq_1.69a}
\mathbb{I}(g)=N\delta(g),\quad \delta(x):=\delta_{xe}=1 \quad {\rm if} \quad x=e, 
\quad \delta(x)=0 \quad {\rm if} \quad x\neq e,
\end{equation}
obviously, $e\in G$ denotes here the group identity. The $\delta$-type convolution identity does exist for any discrete group; it is just $\delta$ itself for finite groups.

Obviously, $\delta$ is identical with the sum of idempotents $\varepsilon(\alpha)$,
\begin{equation}\label{eq_1.70}
\delta=\sum_{\alpha\in \Omega}\varepsilon(\alpha).
\end{equation}
If $G$ is a continuous group, this expression is a divergent series and there is no identity in group algebra.
It does exist in any (in general non-minimal) two-sided ideal obtained from (\ref{eq_1.70}) when the summation
is extended over a finite subset of $\Omega$. The procedure of the formally introduced identity in a 
one-dimensionally extended group algebra would be rather artificial, nevertheless. If $G$ is a Lie group, it is much 
more natural to introduce the "identity" represented by the Dirac-delta distribution. The more so some derivatives of "delta" distribution represent some very important physical quantities.

Let us stress that the minimal two-sided Hermitian idempotents $\varepsilon(\alpha)$ span the center of group algebra;
more precisely, they form a complete system in the subspace of convolution-central functions. All such central functions
are constant on the classes of conjugate elements, i.e., on the orbits of inner automorphisms of $G$. If we admit the 
unit element of group algebra as represented by the delta Dirac distribution, then formally this $\delta$ is given by
the series (\ref{eq_1.70}). As a function series it is divergent, however the limit does exist in the distribution sense.
And, as a functional on the appropriate function space, $\delta$ assigns to any function $F:G\rightarrow \mathbb{C}$
its value at the unit element $e$ of $G$,
\begin{equation}\label{eq_1.71}
\langle \delta, f\rangle=f(e).
\end{equation} 

It is shown that the set of minimal two-sided ideals is identical with the set $\Omega$ of unitary irreducible 
representations of $G$, pairwise non-equivalent ones. More precisely, $\Omega$ is the set of equivalence classes of unitary
irreducible representations. Due to the compactness of $G$, all those representations are finite dimensional.
Let $D(\alpha):G \rightarrow {\rm U}(n(\alpha))\subset {\rm GL}(n(\alpha),\mathbb{C})$ denote the $\alpha$-th unitary 
irreducible representation, more precisely, some representant of the corresponding class. Obviously, $n(\alpha)$
is the dimension of the corresponding representation space $\mathbb{C}^{n(\alpha)}$. All these representation 
spaces are assumed to be unitary in the sense of the standard scalar products,
\begin{equation}\label{eq_1.72}
\langle u,v\rangle=\sum_{k=1}^{n(\alpha)}\overline{u^{a}}v^{a}=\delta_{ab }\overline{u^{a}}v^{b}.
\end{equation} 

The minimal two-sided ideals $M(\alpha)$ are spanned by the matrix elements of the representations $\alpha$,
$D(\alpha)_{ij}$. Moreover, one can show that after appropriate normalization the functions $D(\alpha)_{ij}$
form the canonical basis, more precisely, the canonical complete system of the $H^{+}$-algebra $L^{2}(G)$. This 
follows from the following properties of matrix elements, well-known from the representation theory:
\begin{eqnarray}
D(\alpha)_{ij}*D(\alpha)_{kl}&=&\frac{1}{n(\alpha)}\delta_{jk}D(\alpha)_{il},
\label{eq_1.73a}\\
D(\alpha)_{ij}*D(\beta)_{rs}&=&0 \qquad {\rm if} \quad \beta\neq \alpha,\label{eq_1.73b}\\
\left(D(\alpha)_{ij},D(\alpha)_{kl}\right)&=&\frac{1}{n(\alpha)}\delta_{ik}\delta_{jl},
\label{eq_1.73c}\\
\left(D(\alpha)_{ij},D(\beta)_{rs}\right)&=&0 \qquad {\rm if} \quad \beta\neq \alpha.\label{eq_1.73d}
\end{eqnarray} 
These equations are valid when the Haar measure on $G$ is normed to unity,
\begin{equation}\label{eq_1.74}
\mu(G)=\int_{G}d\mu=1.
\end{equation} 
If other normalization is fixed, then on the right-hand sides of (\ref{eq_1.73a})--(\ref{eq_1.73d}) the volume of $G$,
$\mu(G)$, appears as a factor.

Let $\chi(\alpha)$ and $\varepsilon(\alpha)$ denote the character of $D(\alpha)$
and the corresponding trace of 
$\varepsilon(\alpha)=n(\alpha)D(\alpha)$ respectively:
\begin{equation}\label{eq_1.75}
\chi(\alpha)=\sum_{i}D(\alpha)_{ii}, \qquad \varepsilon(\alpha)=\sum_{i}\varepsilon(\alpha)_{ii}=n(\alpha)\chi(\alpha).
\end{equation} 
Then, obviously,
\begin{eqnarray}
\chi(\alpha)*\chi(\alpha)=\frac{1}{n(\alpha)}\chi(\alpha), &\quad& \chi(\alpha)*\chi(\beta)=0 \quad  {\rm if} \quad \alpha\neq \beta,
\label{eq_1.76a}\\
(\chi(\alpha),\chi(\alpha))=1, &\quad& (\chi(\alpha),\chi(\beta))=0 \quad {\rm if} \quad \alpha\neq \beta,\label{eq_1.76b}
\end{eqnarray}
and similarly
\begin{eqnarray}
\varepsilon(\alpha)*\varepsilon(\alpha)=\varepsilon(\alpha), &\quad& \varepsilon(\alpha)*\varepsilon(\beta)=0 \quad  {\rm if} \quad \alpha\neq \beta, 
\label{eq_1.77a}\\
(\varepsilon(\alpha),\varepsilon(\alpha))=n^{2}(\alpha), &\quad& (\varepsilon(\alpha),\varepsilon(\beta))=0 \quad  {\rm if} \quad \alpha\neq \beta.\label{eq_1.77b}
\end{eqnarray}
The above properties tell us that the canonical basis is given by functions:
\begin{equation}\label{eq_1.78}
\varepsilon(\alpha)_{ij}=n(\alpha)D(\alpha)_{ij}.
\end{equation}
They really satisfy all above-quoted structural properties of canonical bases in $H^{+}$-algebras,
\begin{eqnarray}
\varepsilon(\alpha)_{ij}*\varepsilon(\alpha)_{kl}&=&
\delta_{jk}\varepsilon(\alpha)_{il},
\label{eq_1.79a}\\
\varepsilon(\alpha)_{ij}*\varepsilon(\beta)_{rs}&=&0 \qquad  {\rm if} \quad  \beta\neq \alpha,
\label{eq_1.79b}\\
(\varepsilon(\alpha)_{ij},\varepsilon(\alpha)_{kl})&=&
\delta_{ik}\delta_{jl}n(\alpha),
\label{eq_1.79c}\\
(\varepsilon(\alpha)_{ij},\varepsilon(\beta)_{rs})&=&0 \qquad  {\rm if} \quad  \beta\neq \alpha,
\label{eq_1.79d}\\
\varepsilon(\alpha)_{ij}{}^{+}&=&\varepsilon(\alpha)_{ji},
\label{eq_1.79e}\\
{\rm Tr}\varepsilon(\alpha)_{ij}&=&\delta_{ij}.\label{eq_1.79f}
\end{eqnarray}
Let us stress that the above symbols "+" and "${\rm Tr}$" are used in the sense of (\ref{eq_1.57}), (\ref{eq_1.58}) and do not
concern directly the operations performed on indices $i$, $j$. However, they concern them indirectly. Namely, every function $F\in L^{2}(G)$ may be expanded as a function series with respect to the above complete system:
\begin{equation}\label{eq_1.80}
F=\sum_{\alpha\in\Omega,\ n,m=1,\ldots,n(\alpha)}F(\alpha)_{nm}\varepsilon(\alpha)_{nm}.
\end{equation}
Let us describe relationships (\ref{eq_1.79a})--(\ref{eq_1.79f}) in terms of the coefficients used here; for binary operations we analogously
expand the other function,
\begin{equation}\label{eq_1.81}
G=\sum_{\alpha\in\Omega,\ n,m=1,\ldots,n(\alpha)}G(\alpha)_{nm}\varepsilon(\alpha)_{nm}.
\end{equation}
It follows from the above rules (\ref{eq_1.79a})--(\ref{eq_1.79f}) that the convolution of $F$, $G$ is represented by the system of matrices
$\left(F(\alpha) G(\alpha)\right)_{nm}=\sum_{k}F(\alpha)_{nk}G(\alpha)_{km}$
\begin{equation}\label{eq_1.82}
F*G=\sum_{\alpha\in\Omega,\ n,m=1,\ldots,n(\alpha)}(F(\alpha) G(\alpha))_{nm}\varepsilon(\alpha)_{nm}.
\end{equation}

Similarly, the Hermitian conjugate and the ${\rm Tr}$-functional are represented by the usual matrix Hermitian conjugation and trace:
\begin{eqnarray}
F^{+}&=&\sum_{\alpha\in\Omega,\ n,m=1,\ldots,n(\alpha)}(F(\alpha)^{+} )_{nm}\varepsilon(\alpha)_{nm},\label{eq_1.83a}\\
{\rm Tr}\ F&=&\sum_{\alpha\in\Omega,\ n,m=1,\ldots,n(\alpha)}{\rm Tr}F(\alpha),\label{eq_1.83b}
\end{eqnarray}
in particular, for the scalar product we have that
\begin{equation}\label{eq_1.84}
(F,G)=\sum_{\alpha\in\Omega,\ n,m=1,\ldots,n(\alpha)}{\rm Tr}\left(F(\alpha)^{+}G(\alpha)\right)n(\alpha).
\end{equation}
This is just the explanation of apparently strange definitions of those operations.

In a sense, the group algebra over $G$ may be considered as an arena for some type of the algebraic,
operator-type formulation of quantum mechanics \cite{9}. We are given the associative convolution-product and all necessary
equipment of $H^{+}$-algebra. So, we have everything that is necessary for the algebraic formulation of quantum mechanics.
Physical quantities are $+$-self-adjoint elements of the group algebra, $A^{+}=A$. Density operators are self-adjoint elements $\rho=\rho^{+}$ satisfying in addition the normalization condition,
\begin{equation}\label{eq_1.85}
{\rm Tr}\ \rho=1,
\end{equation}
and the positive-definiteness condition,
\begin{equation}\label{eq_1.86}
\left(\rho, A^{+}*A\right)>0
\end{equation}
for any element $A$ of the group algebra. Pure states are described by Hermitian idempotents; thus, 
in addition to the above conditions the following must hold:
\begin{equation}\label{eq_1.87}
\rho\rho=\rho,\qquad \rho^{+}=\rho.
\end{equation}
Expectation value of the physical quantity $A=A^{+}$ on the state $\rho$ is given by
\begin{equation}\label{eq_1.88}
\langle A\rangle_{\rho}={\rm Tr}\left(A\rho\right)=\left(A^{+},\rho\right)=(A,\rho).
\end{equation}

If $\rho_{0}$ is some pure state, then the probability that the measurement performed on the general
state $\rho$ will detect the state $\rho_{0}$ is given by
\begin{equation}\label{eq_1.89}
{\rm Tr}\left(\rho\rho_{0}\right)=\left(\rho,\rho_{0}\right).
\end{equation}

Statistical interpretation may be also assigned to non-normalized states $\rho$, i.e., ones 
non-satisfying (\ref{eq_1.85}). Then one can say only about relative probabilities.
However, there are yet no wave functions and no superposition principle. It is a good thing to have
also some space of wave functions. The most natural candidates are $L^{2}$-spaces on the group 
$G$ and its homogeneous spaces. Before going any further in this direction one should quote
some comments concerning invariance problems.

Everything above was based on the convolution product (\ref{eq_1.54}). It is evidently associative,
\begin{equation}\label{eq_1.90}
(F*G)*H=F*(G*H).
\end{equation}
It is so for any $L^{1}(G)$-functions on any locally compact topological group $G$. The group structure of $G$
brings about the question concerning the $G$-invariance of the convolution, in any case the question concerning
the sense of such invariance. On the group manifold $G$ the group $G$ itself acts through three natural transformation
groups: Left translations, right translations and inner automorphisms. These actions are given respectively as follows:
\begin{eqnarray}
x\mapsto L_{g}(x):=gx, &\quad& x\mapsto xg:=R_{g}(x)=xg,
\label{eq_1.91a}\\
x\mapsto A_{g}(x):=gxg^{-1}, &\quad& A_{g}=L_{g} \circ R_{g^{-1}}= R_{g^{-1}}\circ L_{g}.\label{eq_1.91b}
\end{eqnarray}
With this convention, $g \mapsto L_{g}$, $g \mapsto A_{g}$, $g\mapsto R_{g}$ are respectively two realizations and anti-realization of the group $G$:
\begin{equation}\label{eq_1.92}
L_{g_{1}g_{2}}=L_{g_{1}}\circ L_{g_{2}}, \qquad A_{g_{1}g_{2}}=A_{g_{1}}\circ A_{g_{2}}, \qquad 
R_{g_{1}g_{2}}=R_{g_{2}}\circ R_{g_{1}}.
\end{equation}
Obviously, substituting $g^{-1}$ instead of $g$ we replace realizations by anti-realiza\-tions and conversely.

The above operations induce the pointwise actions on functions on $G$, namely
\begin{eqnarray}
L[g]f:=f\circ L_{g^{-1}}, &\quad&  R[g]f:=f\circ R_{g^{-1}},\label{eq_1.93a}\\
A[g]f:=f\circ A_{g^{-1}}, &\quad&  A[g]=L[g]R\left[g^{-1}\right]=R\left[g^{-1}\right]L[g].\label{eq_1.93b}
\end{eqnarray}
Due to the replacing of $g$ by $g^{-1}$, we obtain in this way two linear representations and 
anti-representation of $G$ in function spaces on $G$ itself,
\begin{equation}\label{eq_1.94}
L[g_{1}g_{2}]=L[g_{1}]L[g_{2}], \quad A[g_{1}g_{2}]=A[g_{1}]A[g_{2}], \quad R[g_{1}g_{2}]=R[g_{2}]R[g_{1}].
\end{equation}

All these transformations preserve the spaces $L^{1}(G)$, $L^{2}(G)$; they preserve also the scalar products and the 
corresponding statistical statements concerning measurements. However, the left and right regular translations, $L[g]$,
$R[g]$ do not preserve the convolution; unlike this, internal automorphisms do preserve this algebraic structure. Indeed,
\begin{eqnarray}
L[g](F*G)=(L[g]F)*G&\neq& (L[g]F)*(L[g]G),\label{eq_1.95a}\\
R[g](F*G)=F*(R[g]G)&\neq& (R[g]F)*(R[g]G),\label{eq_1.95b}\\
A[g](F*G)&=&(A[g]F)* (A[g]G).\label{eq_1.95c}
\end{eqnarray}
Physically the associative product has to do with spectra, eigenvalues and eigenstates. This is just
this part of physical statements for which the left and right regular translations in $G$ are not physical
automorphisms in the $H^{+}$-algebraic formulation of quantum mechanics. Concerning the connection with
spectra and eigenproblems: in an algebraic formulation, including the $H^{+}$-algebraic one, the number $\lambda$
does belong to the spectrum of the function $F$ if the convolution inverse of $\left(F-\lambda\delta\right)$ does
not exist, i.e., if there is no function $H$ satisfying
\begin{equation}\label{eq_1.96}
H*\left(F-\lambda\delta\right)=\delta.
\end{equation}
This is easily expressible in terms of the expansion (\ref{eq_1.80}), namely $\lambda\in {\rm Sp}\: F$ if and only if there
exists $\alpha\in \Omega$ such that
\begin{equation}\label{eq_1.97}
\det\left(F(\alpha)-\lambda{\rm I}_{n(\alpha)}\right)=0.
\end{equation}
All this is equivalent to the statement that there exists such a "density matrix" $\rho\in L^{2}(G)$ that the following "eigenequation" holds:
\begin{equation}\label{eq_1.98}
A*\rho=\lambda\rho;
\end{equation}
when $A^{+}=A$, this is equivalent to the right-hand-side "eigenequation":
\begin{equation}\label{eq_1.99}
\rho*A=\lambda\rho.
\end{equation}
These physically interpretable statements are based on the associative convolution product, therefore, the left and right regular translations, which do not preserve it, are not physical symmetries of the quantum-mechanical formulation 
based on the group algebra of $G$. Some light is shed on such problems when convolutions are interpreted as linear shells of regular translations. And at the same time some natural link is established then with the concepts of wave functions, superpositions, etc.

Let us follow one of finite-dimensional patterns outlined above. Namely, we begin with the linear space $H$ of wave functions on $G$, in principle the Hilbert space $L^{2}\left(G,dg\right)$, although in practical problems the Hilbert space
language is often too narrow, e.g., one must admit distributions or non-normalizable wave functions (in the non-compact case). The following sets are relevant for quantum theory: the Banach algebra $B(H)$ of bounded linear operators on $H$
and $H^{+}$-algebraic structures in appropriate subspaces of $B(H)$. Of course, in practical problems some non-bounded operators, elements of $L(H)$ are admissible and, when properly and carefully treated, just desirable. The point is that some very important physical quantities, e.g., momenta, angular momenta and so on, are represented by differential operators, 
obviously, non-bounded ones. 

However, let us begin with bounded operators describing $G$-symmetries, namely (\ref{eq_1.93a}), (\ref{eq_1.93b}),
$L[g]$, $R[g]$, $A[g]=L[g]R\left[g^{-1}\right]=R\left[g^{-1}\right]L[g]$. Obviously, these operators are unitary in $L^{2}\left(G,dg\right)$. Being unitary, they are obviously bounded. The linear shell of the family of operators $\left\{L[g]:g\in G\right\}$ is just the group algebra
of $G$. Namely, if we take two functions $F,H\in L^{1}(G)$ and the corresponding linear operators
\begin{equation}\label{eq_1.100}
L\{F\}=\int F(g)L[g]dg, \qquad L\{H\}=\int H(g)L[g]dg,
\end{equation}
then it may be easily shown that
\begin{equation}\label{eq_1.101}
L\{H\}L\{F\}=L\{H*F\}.
\end{equation}
Obviously (\ref{eq_1.100}) is a rather symbolic way of writing; this formula is meant in the following sense:
\begin{equation}\label{eq_1.100a}
L\{F\}f=\int F(g)L[g]fdg=\left(F*f\right).
\end{equation}

Therefore, all operations performed on operators of the type $L\{F\}$ are represented by the corresponding, described above, operations on functions $F$ as elements of the Lie algebra over $G$. Let us stress that the operators of the form 
(\ref{eq_1.100}) are very special, albeit important elements of $L(H)$. Let us formally substitute for $F$ in (\ref{eq_1.100}) the "delta distribution" $\delta_{h}$ concentrated at $h\in G$, i.e., symbolically, 
\begin{equation}\label{eq_1.102}
\delta_{h}(g)=\delta\left(gh^{-1}\right)=\delta\left(hg^{-1}\right).
\end{equation}
Then, obviously, 
\begin{equation}\label{eq_1.103}
 L\{\delta_{h}\}=L[h], \quad \delta_{h}*f=L[h]f,
\end{equation}
i.e., the formal convolution with $\delta_{h}$ is the $h$-translation of $f$. In particular, $\delta_{e}=\delta$ is the convolution identity. Something similar may be done with the right translations; one obtains then another family of linear operators acting on wave functions. Namely, let us take again the linear shell of right regular translations, in particular the operators:
\begin{equation}\label{eq_1.104}
 R\{F\}=\int F(g)R[g]dg, \qquad R\{H\}=\int H(g)R[g]dg,
\end{equation}
obviously, with the definition analogues to (\ref{eq_1.100a}), thus,
\begin{equation}\label{eq_1.104a}
R\{F\}f=f*F.
\end{equation}
And again after simple calculations we obtain the following superposition rule:
\begin{equation}\label{eq_1.105}
R\{F\}R\{H\}=R\{H*F\}.
\end{equation}
Unlike the representation rule (\ref{eq_1.101}), this is antirepresentation of the convolution group algebra on $G$ into the algebra of all operators acting on "wave functions" on $B$, in particular, on $L^{1}(G)$, $L^{2}(G)$. To obtain the representation property also for the $R$-objects, one should define them in the "transposed" way:
\begin{eqnarray}
R^{T}\{F\}&:=&\int F^{T}(g)R[g]dg=\int F\left(g^{-1}\right)R[g]dg=\int F(g)R\left[g^{-1}\right]dg,\qquad
\label{eq_1.104b1}\\
F^{T}(g)&:=&F\left(g^{-1}\right).\label{eq_1.104b2}
\end{eqnarray}
Then, obviously, 
\begin{equation}\label{eq_1.105a}
 R^{T}\{F\}R^{T}\{H\}=R^{T}\{F*H\}.
\end{equation}

The transformation rules (\ref{eq_1.95a}), (\ref{eq_1.95b}) and the representation rules (\ref{eq_1.101}), (\ref{eq_1.105}) tell us that the convolution is not invariant under regular translations, i.e., convolution of translates differs from the translate of convolution. Nevertheless, this is a very peculiar non-invariance and something is invariant, in a sense. Namely, the left-translated convolution is identical with the convolution in which the left factor is left-translated and the right one kept unchanged. And conversely, the right-translated convolution is the convolution in which the right factor is right-translated (but only this one). 

Concerning translational non-invariance of the convolution, let us notice that $F*H$ may be symbolically expressed with the use of the Dirac distribution:
\begin{equation}\label{eq_1.106}
\left(F*H\right)(g)=\int \delta\left(x^{-1}gy^{-1}\right)F(x)H(y)dxdy=\int \delta\left(yg^{-1}x\right)F(x)H(y)dxdy.
\end{equation}

Let us write down a binary multilinear operation on functions on $G$ in the integral form, maybe symbolic one:
\begin{equation}\label{eq_1.107}
\left(F\perp H\right)(g)=\int K(g;x,y)F(x)H(y)dxdy.
\end{equation}
This operation is invariant under right or left translations, i.e., respectively the following holds:
\begin{eqnarray}
\left(F\perp H\right)(gh)&=&\int K(g;x,y)F(xh)H(yh)dxdy,\label{eq_1.108}\\
\left(F\perp H\right)(hg)&=&\int K(g;x,y)F(hx)H(hy)dxdy,\label{eq_1.109}
\end{eqnarray}
when
\begin{equation}\label{eq_1.110}
K(g;x,y)=K_{r}\left(gx^{-1},gy^{-1}\right), \qquad K(g;x,y)=K_{l}\left(x^{-1}g,y^{-1}g\right)
\end{equation}
respectively for (\ref{eq_1.108}) and (\ref{eq_1.109}). Obviously, (\ref{eq_1.106}), $K(g;x,y)=\delta\left(x^{-1}gy^{-1}\right)$ does not satisfy any of conditions (\ref{eq_1.108}), (\ref{eq_1.109}). Moreover, if $G$ is non-Abelian, those two conditions are rather incompatible.

Let us consider also the total linear shell of translation operators. Let us take a function $F:G\times G \rightarrow \mathbb{C}$ and construct the operator
\begin{equation}\label{eq_1.111}
T_{t}\{F\}:=\int F\left(g_{1},g_{2}\right)L[g_{1}]R\left[g_{2}^{-1}\right]dg_{1}dg_{2}.
\end{equation}
One can show that multiplication of such operators results in convolution of functions on the direct product $G \times G$,
\begin{equation}\label{eq_1.112}
T_{t}\{F\}T_{t}\{H\}=T_{t}\{F*H\},
\end{equation}
where
\begin{equation}\label{eq_1.113}
\left(F*H\right)\left(g_{1},g_{2}\right)=\int F(h_{1},h_{2})H\left(h_{1}{}^{-1}g_{1},h_{2}{}^{-1}g_{2}\right)dh_{1}dh_{2}.
\end{equation}

Obviously, one could also proceed like in (\ref{eq_1.104}), namely, define
\begin{equation}\label{eq_1.114}
T\{F\}=\int F(g_{1},g_{2})L[g_{1}]R[g_{2}]dg_{1}dg_{2}.
\end{equation}
Then the convolution in the second argument is "transposed"; by that we mean the operation
\begin{equation}\label{eq_1.115}
f*^{t}g:=g*f,
\end{equation}
thus,
\begin{equation}\label{eq_1.116}
T\{F\}T\{H\}=T\{F\left(*^{t}\right)H\}
\end{equation}
where
\begin{equation}\label{eq_1.117}
\left(F\left(*^{t}\right)H\right)(g_{1},g_{2})=\int F(h_{1},h_{2})H\left(h_{1}^{-1}g_{1},g_{2}h_{2}^{-1}\right)dh_{1}dh_{2}.
\end{equation}
Obviously, the difference between (\ref{eq_1.114}), (\ref{eq_1.116}), (\ref{eq_1.117}) and respectively (\ref{eq_1.111}), (\ref{eq_1.112}), (\ref{eq_1.113}) is of a rather cosmetical nature.

Operators of the convolution form (\ref{eq_1.111})/(\ref{eq_1.114}) are very special linear operations acting on the wave functions $\Psi:G\rightarrow \mathbb{C}$. They are "smeared out" in $G$, essentially non-local, if $F$ is a "usual", "good" function. To obtain very important operators of geometrically distinguished physical quantities or unitary operators of left and right regular translations one must use distributions. Let us mention, e.g., the obvious, trivial examples:
\begin{equation}\label{eq_1.118}
L[h]=L\{\delta_{h}\}, \qquad R[h]=R\{\delta_{h}\}, \qquad L[h]R[k]=F\{\delta_{(h,k)}\},
\end{equation}
where, let us repeat,
\begin{equation}\label{eq_1.119}
\delta_{h}(g)=\delta\left(gh^{-1}\right), \quad \delta_{h,k}\left(g_{1},g_{2}\right)=
\delta\left(g_{1}h^{-1},g_{2}k^{-1}\right)=\delta_{h}(g_{1})\delta_{k}(g_{2}).
\end{equation}

A more detailed analysis, including the description of important physical quantities, may be performed only when one deals with Lie groups and makes use of their differential and analytical structure. Here we stress only the fact that when the algebraic scheme of group algebras is used, then the regular translations fail to be automorphisms of the theory. Physical automorphisms are given by the operators $A[g]=L[g]R\left[g^{-1}\right]=R\left[g^{-1}\right]L[g]$. Therefore, the minimal two-sided ideals $M(\alpha)$ must be further decomposed into direct sums of minimal subspaces invariant under inner automorphisms. This leads to operator algebras invariant under unitary similarity transformations acting in the space $L^{2}(G)$ of wave functions.

Let us stress that the Hilbert space operations in $L^{2}(G)$ as the space of wave functions are compatible with the $H^{+}$-algebra operations in the sense that 
\begin{equation}\label{eq_1.120}
L\{F\}^{+}=L\left\{F^{+}\right\}, \qquad R\{F\}^{+}=R\left\{F^{+}\right\}, \qquad T\{F\}^{+}=T\left\{F^{+}\right\},
\end{equation}
where the involutions on the right-hand sides of these equations are meant in the sense of (\ref{eq_1.57}), i.e.,
\begin{equation}\label{eq_1.121}
F^{+}(g)=\overline{F\left(g^{-1}\right)}, \qquad F^{+}\left(g_{1},g_{2}\right)=\overline{F\left(g_{1}^{-1},g_{2}^{-1}\right)}.
\end{equation}
In particular, the mentioned operators are Hermitian if and only if the corresponding functions are involution-invariant.

Similarly, the operators of convolution are unitary, 
\begin{eqnarray}
L\left\{ F\right\} ^{+}L\left\{ F\right\} &=&{\rm Id}_{L^{2}(G)},\label{eq_1.122a} \\ 
R\left\{ F\right\} ^{+}R\left\{ F\right\} &=&{\rm Id}_{L^{2}(G)},\label{eq_1.122b}\\
T\left\{ H\right\} ^{+}T\left\{ H\right\} &=&{\rm Id}_{L^{2}(G\times G)},
\label{eq_1.122c}
\end{eqnarray}
if and only if 
\begin{equation}
F^{+}* F=\delta_{G},\quad H^{+}* H=\delta_{G\times G}.\label{eq_1.123}
\end{equation}

Obviously, all translation operators (\ref{eq_1.93a}), (\ref{eq_1.93b}), (\ref{eq_1.118}), (\ref{eq_1.119}) are unitary in $L^{2}(G)$; 
this follows from the
invariance of the Haar measure. They preserve the scalar product of
wave functions on $G$, and automatically they preserve the scalar
product in the $H^{+}$-algebra $L^{2}(G)$. It is interesting to
stress again how they are represented in the algebra $L(L^{2}(G))$
of all operators in $L^{2}(G)$, and first of all, in the algebra
of bounded operators $B(L^{2}(G))$. Obviously, any invertible operator
$F$ in $L^{2}(G)$ acts in $L(L^{2}(G))$, $B(L^{2}(G))$ through
the similarity transformations: 
\begin{equation}
A\rightarrow FAF^{-1}.\label{eq_1.124}
\end{equation}
This concerns in particular unitary operators, like regular translations.
It is interesting to see how they act in the linear shell of translations,
i.e., in the algebras of convolution-type operators. One can easily
show that:
\begin{eqnarray}
L[g]L\{F\}L[g]^{-1} & =&L\{A[g]F\},\label{eq_1.125a}\\
R[g]L\{F\}R[g]^{-1} & =&L\{F\},\label{eq_1.125b}\\
L[g]R\{F\}L[g]^{-1} & =&R\{F\},\label{eq_1.125c}\\
R[g]R\{F\}R[g]^{-1} & =&R\{A[g^{-1}]F\},\label{eq_1.125d}
\end{eqnarray}
or, better to express,
\begin{equation}
R[g^{-1}]R\{F\}R[g^{-1}]^{-1}=R\{A[g]F\}.\label{eq_1.126}
\end{equation}
Therefore, 
\begin{equation}
L[g_{1}]R[g_{2}]T\{F\}\left(L[g_{1}]R[g_{2}]\right)^{-1}=T\left\{ F\circ\left(A[g_{1}]\times A[g_{2}\!^{-1}]\right)\right\},\label{eq_1.127}
\end{equation}
or equivalently, 
\begin{equation}
L[g_{1}]R[g_{2}]T_{t}\{F\}\left(L[g_{1}]R[g_{2}]\right)^{-1}=T_{t}\left\{ F\circ\left(A[g_{1}]\times A[g_{2}]\right)\right\},\label{eq_1.128}
\end{equation}

Obviously, the Cartesian product of mappings $A:X\rightarrow U$,
$B:Y\rightarrow V$, denoted by $A\times B:X\times Y\rightarrow U\times V$,
is meant in the usual sense, i.e., 
$\left(A\times B\right)(x,y)=\left(A(x),B(y)\right)$. 
The message of these formulas is that regular translations acting
on wave functions on $G$ are represented in the group algebras over
$G$ or $G\times G$ by inner automorphisms. Because of this, classification
of states and physical quantities in group algebras over $G$ and
$G\times G$ is based on the analysis of minimal subspaces invariant
under inner automorphisms. One point must be stressed: the operators
of the form $T\{F\}$ are not the most general operators acting in
$L^{2}(G)$. The position operators, more precisely, the operators
of pointwise multiplication of wave functions by functions on $G$
do not belong to this class. Nevertheless, if $G$ is non-Abelian,
then some (not all!) position-like quantities are implicitly present
in $T\{f\}$-type operators.

The main peculiarity of convolution-type operators and their singular special cases like the regular translations and automorphisms (\ref{eq_1.93a}), (\ref{eq_1.93b}), (\ref{eq_1.100}), (\ref{eq_1.104}), (\ref{eq_1.111}), (\ref{eq_1.113}) is that they preserve separately all subspaces/ideals $M(\alpha)$. 

It is not the case with the position-like operators; they mix various
subspaces (ideals) with each others. The points is interesting in
itself, because it has to do with the relationship between two algebraic
structures in function spaces over $G$. One of them is just the structure
of the associative algebra under convolution in $L^{1}(G)$. It is
non-commutative unless $G$ is Abelian and has no literally meant
unity unless $G$ is discrete. The other one is the structure of commutative
associative algebra under the pointwise multiplication of functions.
Obviously, this is the unital algebra with the unity given by the
constant function taking the value $1$ all over $G$. In general
there are some subtle points concerning the underlying sets of algebras
and their mutual relationships. Obviously, the convolution algebra
is defined in $L^{1}(G)$, whereas the pointwise multiplicative
algebra is defined in the set of all globally defined functions on
$G$. No comments are necessary, and the both underlying sets coincide,
only when $G$ is finite 

The pointwise products of matrix elements of representations $D(\alpha)$,
$D(\varrho)$ are expanded with respect to the orthogonal systems
of functions $D(\varkappa)_{kl}$ according to the rule: 
\begin{equation}
D(\alpha)_{ab}D(\varrho)_{rs}=\underset{\varkappa,k,l}{\sum}\left(\alpha\varrho ar|\varkappa k\right)\left(\alpha\varrho bs|\varkappa l\right)D(\varkappa)_{kl},\label{eq_1.129}
\end{equation}
where $\left(\alpha\varrho ar|\varkappa k\right)$ and so on are Clebsch-Gordan
coefficients for the group $G$ \cite{4,6,7}. Summation over $\varkappa$ is extended
over an appropriate range depending on $(\alpha,\varrho)$, and obviously,
for the fixed $\kappa$, the range of $k$, $l$ is given by the set of
naturals $1,\ldots,n(\varkappa)$. To be more precise, $k$, $l$ run over
some $n(\varkappa)$-element set. In the theory of angular momentum,
when $G={\rm SU}(2)$, it is convenient to use the convention: $\varkappa=2j+1$,
where $j$ runs over the set of non-negative integers and half-integers,
and $k$, $l$ run over the range $-j,-j+1,...,j-1,j$, jumping by one.
Obviously, (\ref{eq_1.78}) implies that 
\begin{equation}
\varepsilon(\alpha)_{ab}\varepsilon(\varrho)_{rs}=
\underset{\varkappa,k,l}{\sum}\frac{n(\alpha)n(\varrho)}{n(\varkappa)}
\left(\alpha\varrho ar|\varkappa k\right)\left(\alpha\varrho bs|\varkappa l\right)\varepsilon(\varkappa)_{kl}.\label{eq_1.130}
\end{equation}
The coefficients at $\varepsilon(\varkappa)_{kl}$ are structure constants of the commutative algebra of pointwise multiplication
with respect to the canonical basis/complete system. They are bilinear
in Clebsch-Gordon coefficients. The latter ones are meant in the usual
sense of the procedure:
\begin{itemize}
\item[$(i)$] Take two irreducible representations of $G$, $D(\alpha)$, $D(\varrho)$ acting respectively in $\mathbb{C}^{n(\alpha)}$, $\mathbb{C}^{n(\varrho)},
$\begin{eqnarray}
\left(D(\alpha)(g)u(\alpha)\right)_{a}&= & \underset{b}{\sum}D(\alpha)(g)_{ab}u(\alpha)_{b},\label{eq_1.131a}\\
\left(D(\alpha)(g)u(\varrho)\right)_{r}&= & \underset{s}{\sum}D(\alpha)(g)_{rs}u(\varrho)_{s}.
\label{eq_1.131b}
\end{eqnarray}

\item[$(ii)$] Take the tensor product of those representations, 
\begin{equation}\label{eq_1.132a}
D(\alpha)\otimes D(\varrho):\qquad G\times G\rightarrow L(\mathbb{C}^{n(\alpha)}\otimes\mathbb{C}^{n(\varrho)})\simeq L(\mathbb{C}^{n(\alpha)n(\varrho)}),
\end{equation}
given by
\begin{equation}\label{eq_1.132b}
\left(\left(D(\alpha)(g_{1})\otimes D(\varrho)(g_{2})\right)t(\alpha,\varrho)\right)_{ar} =
\underset{bs}{\sum}D(\alpha)(g_{1})_{ab}
D(\varrho)(g_{2})_{rs}t(\alpha,\varrho)_{bs}.
\end{equation}
This representation is irreducible, if, as assumed, $D(\alpha)$, $D(\varrho)$
are irreducible. 

\item[$(iii)$] Take the direct product of $D(\alpha)$, $D(\varrho)$, i.e., restrict $D(\alpha)\otimes D(\varrho)$ to the diagonal $\left\{(g,g):g\in G\right\}\subset G\times G$. One obtains some representation $D(\alpha)\times D(\varrho)$ of $G$ in $\mathbb{C}^{n(\alpha)}\otimes\mathbb{C}^{n(\beta)}\simeq
\mathbb{C}^{n(\alpha)n(\beta)}$. 

In general, this representation is reducible and equivalent to the
direct sum of some irreducible representations,
\begin{equation}
\underset{\varkappa\in\Omega(\alpha,\varrho)}{\bigoplus}D(\varkappa);
\label{eq_1.133}\\
\end{equation}
the direct sum performed over some subset of labels, $\Omega\left(\alpha,\varrho\right)\subset\Omega$.
Obviously, this representation acts in the Cartesian product 
\begin{equation}
\underset{\varkappa\in\Omega\left(\alpha,\varrho\right)}{\times}
\mathbb{C}^{n\left(\varkappa\right)}.\label{eq_1.134}
\end{equation}
Let $U$ denote an equivalence isomorphism of $\mathbb{C}^{n(\alpha)}\otimes\mathbb{C}^{n(\varrho)}$
onto the representation space (\ref{eq_1.134}). Then, by definition,
the Clebsch-Gordon coefficients are given by
\begin{equation}
U\left(u\left(\alpha\right)_{a}\otimes v\left(\varrho\right)_{r}\right)=
\underset{\varkappa,k}{\sum}\left(\alpha\varrho ar|\varkappa k\right)w\left(\varkappa\right)_{k},\label{eq_1.135}
\end{equation}
where $u\left(\alpha\right)_{a}$, $v\left(\varrho\right)_{r}$, $w\left(\varkappa\right)_{k}$
denote basis vectors of the representation spaces for $D\left(\alpha\right)$,
$D\left(\varrho\right)$, $D\left(\varkappa\right)$. When the natural
bases in $\mathbb{C}^{n\left(\alpha\right)}$, $\mathbb{C}^{n\left(\varrho\right)}$,
$\mathbb{C}{}^{n\left(\varkappa\right)}$ are used, then $u\left(\alpha\right)_{a}$,
$v\left(\varrho\right)_{r}$, $w\left(\varkappa\right)_{k}$ may be
reinterpreted as components of the representation vectors $u\left(\alpha\right)$,
$v\left(\varrho\right)$, $w\left(\varkappa\right)$. And then we simply
write instead of (\ref{eq_1.135}) the following formulas:
\begin{equation}
u(\alpha)_{a}v(\varrho)_{r}=\underset{\varkappa,k}{\sum}\left(\alpha\varrho ar|\varkappa k\right)w(\varkappa)_{k}.\label{eq_1.136}
\end{equation}
These are just the implicit definitions of the Clebsch-Gordon coefficients.
\end{itemize}

There are two special cases when all minimal ideals $M(\alpha)$ are
finite-dimen\-sional, i.e., all irreducible unitary representations
$D(\alpha)$ are finite-dimensional. These are when the topological
group $G$ is compact or Abelian; of course, those are non-disjoint
situations. In the Abelian case all $M(\alpha)$ are one-dimensional
and one is dealing with Pontryagin duality \cite{10,Rudin}. The set $\Omega$ of irreducible
unitary representations has the natural structure of a locally compact
Abelian group too, the so-called character group, denoted traditionally
by $\widehat{G}$. The group operation in $\widehat{G}$ is meant
as the pointwise multiplication of functions on $G$. In the other
words, the elements of $\widehat{G}$ are continuous homomorphisms
of $G$ into the group $U(1)=\left\{z\in\mathbb{C}:\left|z\right|=1\right\}$,
the multiplicative group of complex numbers of modulus one. If $G$ is
compact and Abelian, then $\widehat{G}$ is discrete, and the Peter-Weyl
series expansion (\ref{eq_1.80}), (\ref{eq_1.81}) becomes a generalized Fourier
series. If $G$ is non-compact, one obtains generalized Fourier transforms
and direct integrals of family of one-dimensional spaces.

According to the well-known Pontryagin theorem, the dual of $\widehat{G}$,
e.g., the second dual $\widehat{\widehat{G}}$ of $G$, is canonically
isomorphic with $G$ itself \cite{10,Rudin}. This resembles the relationship between
duals of finite-dimensional linear spaces, $\left(V^{*}\right)^{*}\simeq V$.

The Fourier transform $\widehat{\Psi}:\widehat{G}\rightarrow\mathbb{C}$ of $\Psi:G\rightarrow\mathbb{C}$ is defined as follows:
\begin{equation}
\widehat{\Psi}(\chi)=\int\overline{\left\langle \chi|g\right\rangle }\Psi(g)dg=\int\left\langle \chi|g\right\rangle ^{-1}\Psi(g)dg,\label{eq_1.137}
\end{equation}
where $dg$ again denotes the integration element of the Haar measure
on $G$ and $\left\langle\chi|g\right\rangle$ is the evaluation
of $\chi\in\widehat{G}$ on $g\in G$; equivalently, in virtue of
Poincare duality, this is the evaluation of $g\in G\simeq\widehat{\widehat{G}}$
on $\chi\in\widehat{G}$. The inverse formula of \ref{eq_1.137} reads:
\begin{equation}
\Psi(g)=\int\left\langle \chi|g\right\rangle \widehat{\Psi}(\chi)d\chi,\label{eq_1.138}
\end{equation}
where $d\chi$ denotes the element of Haar integration on $\widehat{G}$.
The formulas (\ref{eq_1.137}), (\ref{eq_1.138}) fix the synchronization
between normalizations of measures $dg$, $d\chi$. In principle,
these formulas are meant in the sense of $L^{1}$-spaces over $G$,
$\widehat{G}$, nevertheless, some more or less symbolic expressions
are also admitted for other functions, as shorthands for longer systems
of formulas. First of all, this concerns $\delta$-distributions,
just like in general situation of locally compact $G$.
Of course, the correct definition of distributions and operations on them
must be based on differential concepts, nevertheless, the Dirac distribution
itself (but not its derivatives) may be introduced in principle on
the basis of purely topological concepts, just like in the general
case. Let us notice that 
\begin{equation}
\Psi(g)=\int d\chi\int dh\Psi(h)\left\langle \chi|hg^{-1}\right\rangle.\label{eq_1.139}
\end{equation}
The order of integration here is essential! But, of course, one cannot
resist the temptation to change "illegally" this order and write
symbolically:
\begin{equation}
\Psi(g)=\int dh\,\delta\left(hg^{-1}\right)\Psi(h),\qquad\delta(x)=\int d\chi\left\langle \chi|x\right\rangle.\label{eq_1.140}
\end{equation}
If $G$ is discrete, then $\widehat{G}$ is compact (and conversely)
and the second integral is well defined, namely 
\begin{equation}
\delta(x)=\delta_{xe}
\begin{cases}
1, & \textrm{if }\ x=e,\\
0, & \textrm{if }\ x\neq e,
\end{cases}\label{eq_1.141}
\end{equation}
where, obviously, $e$ is the natural element (identity) of $G$.
Then the first integral is literally true as a summation with the
use of Kronecker delta. But when obeying some rules, we may safely
use the formulas (\ref{eq_1.140}) also in the general case, when they
are formally meaningless. So, we shall always write
\begin{eqnarray}
\delta(g)&=&\int\left\langle \chi|g\right\rangle d\chi=\delta\left(g^{-1}\right), \label{eq_1.142a}\\
\delta(\chi)&=&\int\overline{\left\langle \chi|g\right\rangle}dg=
\int\left\langle \chi|g\right\rangle dg=\delta\left(\chi^{-1}\right),\label{eq_1.142b}\\
\int\delta(g)f(g)\, dg&=&f\left(e(G)\right),\label{eq_1.142c}\\
\int\delta(\chi)k(\chi)\, d\chi&=&
k\left(e(\widehat{G})\right),\label{eq_1.142d}
\end{eqnarray}
and $e(G)$, $e(\widehat{G})$ denote the units in $G$, $\widehat{G}$,
respectively. 

Convolution is defined by the usual formula (\ref{eq_1.54}),
but, obviously, the peculiarity of Abelian groups $G$ is that convolution
is a commutative operation, 
\begin{equation}
F*G=G*F.\label{eq_1.143}
\end{equation}
Obviously, Fourier transforms of convolution are pointwise products
of Fourier transforms, and conversely,
\begin{equation}
\left(F*G\right)^{\wedge}=\widehat{F}\widehat{G}.\label{eq_1.144}
\end{equation}
This is the obvious special case of (\ref{eq_1.82}); $\widehat{F}(\chi)$,
$\widehat{G}(\chi)$ are $1\times 1$ matrices $F(\alpha)$, $G(\alpha)$.

It is clear that just like in the general case, $\delta$-distribution
is the convolution identity, 
\begin{equation}
F*\delta=\delta*F=F.\label{eq_1.145}
\end{equation}

And now, we may be a bit more precise. Namely, let $U\subset\widehat{G}$
be some compact measurable subset of $\widehat{G}$, 
and let $L\left\{U\right\}$ denote
the linear subspace of functions (\ref{eq_1.138}) such that the Fourier
transform $\widehat{\Psi}$ vanishes outside $U$ and is $L^{1}$-class.
Take the function $\delta\{U\}$ given by
\begin{equation}
\delta\{U\}(g):=\underset{U}{\int}\left\langle \chi|g\right\rangle d\chi.\label{eq_1.146}
\end{equation}
It is clear that $\delta\{U\}$ is a convolution identity of the subspace
$L\{U\}$. And now take an increasing sequence of subsets $V_{i}\subset\widehat{G}$ such that:
\begin{equation}\label{eq_1.146a}
V_{i}\supset V_{j}\quad\textrm{for}\quad i>j\,\qquad \underset{i}{\bigcup}\ V_{i}=\widehat{G}.
\end{equation}
It is clear that for any function $F\subset L^{1}(G)$ we have 
\begin{equation}
\underset{i\rightarrow\infty}{\lim}\delta\left\{ V_{i}\right\} *F=F,\label{eq_1.147}
\end{equation}
although the limit of the sequence $\delta\{V_{i}\}$ does not exist
in the usual sense of function sequences. However, it does exist in
an appropriately defined functional sense. So, by abuse of language,
we simply write:
\begin{equation}
\delta=\underset{i\rightarrow\infty}{\lim}\delta\{V_{i}\},\qquad\delta*F=F,
\label{eq_1.148}
\end{equation}
as a shorthand for the rigorous (\ref{eq_1.147}). 

Calculating formally the convolution of $\chi_{1},\chi_{2}\in\widehat{G}$,
we obtain 
\begin{equation}
\left\langle \chi_{1}*\chi_{2}|g\right\rangle =\delta\left(\chi_{1}\chi_{2}\!^{-1}\right)\left\langle \chi_{2}|g\right\rangle =\delta\left(\chi_{1}\chi_{2}\!^{-1}\right)\left\langle \chi_{1}|g\right\rangle,\label{eq_1.149}
\end{equation}
i.e., briefly 
\begin{equation}
\chi_{1}*\chi_{2}=\delta\left(\chi_{1}\chi_{2}\!^{-1}\right)\chi_{2}=
\delta\left(\chi_{1}\chi_{2}\!^{-1}\right)\chi_{1}.\label{eq_1.150}
\end{equation}

If $G$ is compact, i.e., $\widehat{G}$ is discrete, this is the
usual condition for irreducible idempotents (\ref{eq_1.77a}), (\ref{eq_1.77b}). Similarity, we have the orthogonality/normalization condition 
\begin{equation}
\left(\chi_{1},\chi_{2}\right)=\delta\left(\chi_{1}\chi_{2}\!^{-1}\right)=
\begin{cases}
1, & \textrm{if }\ \chi_{1}=\chi_{2},\\
0, & \textrm{if }\ \chi_{1}\neq\chi_{2}.
\end{cases}\label{eq_1.151}
\end{equation}

If $G$ is not compact, i.e., $\widehat{G}$ is not discrete, then
both normalization and idempotence rules (\ref{eq_1.150}), (\ref{eq_1.151})
are meant symbolically, just like the corresponding rules for Dirac
distributions in $\mathbb{R}^{n}$: 
\begin{eqnarray}
\delta_{a}*\delta_{b} & = & \delta(a-b)\delta_{a}=\delta(a-b)\delta_{b},\label{eq_1.152a}\\
\left(\delta_{a},\delta_{b}\right) & = & \delta(a-b).\label{eq_1.152b} \end{eqnarray}

Obviously, $\delta_{a}(x):=\delta\left(x-a\right)$. Incidentally,
(\ref{eq_1.152a}), (\ref{eq_1.152b}) is just the special case of (\ref{eq_1.150}), (\ref{eq_1.151})
when $G=\mathbb{R}^{n}$ and the addition of vectors is meant as a
group operation.

The peculiarity of locally compact Abelian groups is that they offer
some analogies to geometry of the classical phase spaces and some
natural generalization of the Weyl-Wigner-Moyal formalism. Certain
counterparts do exist also in non-Abelian groups, especially compact
ones. However, they are radically different from the structures based
on Abelian groups. And in the non-compact case the analogy rather diffuses.

Finally, let us remind that just like in the classical Fourier analysis,
the Pontryagin Fourier transform is an isometry of $L^{2}(G)$ onto
$L^{2}\left(\widehat{G}\right)$,
\begin{equation}
\int\overline{A(g)}B(g)dg=\int\overline{\widehat{A}}(\chi)B(\chi)d\chi,
\label{eq_1.153}
\end{equation}
in particular, the Plancherel theorem holds 
\begin{equation}
\int\left|A(g)\right|^{2}dg=\int\left|\widehat{A}(\chi)\right|^{2}d\chi.
\label{eq_1.154}
\end{equation}
Compare with the formula (\ref{eq_1.84}) for compact topological groups
and the corresponding expression for the norm $\left\| F\right\| $:
\begin{equation}
\left\|F\right\|^{2}=
\underset{\alpha\in\Omega}{\sum}{\rm Tr}\left(F\left(\alpha\right)^{+}F
\left(\alpha\right)\right)n\left(\alpha\right).\label{eq_1.154a}
\end{equation}

Let us finish this part with some comments concerning physical interpretation.

It is impossible to answer definitely the question: "What
is the most fundamental mathematical structure underlying quantum
mechanics?" There are approaches based on quantum
logics, the usual Hilbert space formulations, operator algebras,
etc. According to certain views \cite{8}, all non-artificial
and viable models, both in quantum and classical mechanics, assume
some groups as fundamental underlying structures. The framework of
group algebra may appear in two different, nevertheless, somehow interrelated,
ways.
\begin{enumerate}
\item The first scheme is one in the spirit of algebraic approaches. Namely,
when some topological group $G$ is assumed, one can simply say: "quantum
mechanics based on $G$ is the group algebra of $G$".
In any case it works on compact groups and locally compact Abelian
ones. Group algebras are particular $H^{+}$-algebras. The important
operations of quantum mechanics are based on structures intrinsically
built into them. Hermitian elements represent physical quantities,
Hermitian and positive ones are quantum states in the sense of density
operators, idempotent ones among them represent pure states. This
is common to all $H^{+}$-algebras. The peculiarity of group algebras
is that besides the convolution operation there exist also another
composition rule, namely, the pointwise product of functions, obviously
associative one as well, and commutative (convolution is non-commutative
if $G$ is non-Abelian). The relationship between them is given by
the Clebsch-Gordan coefficients. This structure is also physically interpretable,
namely, it describes the properties of composed system, e.g., composition
of angular momenta. Pointwise multiplication of functions representing
quantum states describes the direct product of density operators of
subsystems \cite{9,6}.

However, one important structure of quantum mechanics is missing
here, namely, the superposition principle. The point is that wave functions
do not fit this framework directly. Nevertheless, in a sense they
are implicitly present. Namely, group algebra, as any associative
algebra, acts on itself through the left or right regular translations,
\begin{equation}
x\rightarrow a*x,\qquad x\rightarrow x*a.\label{eq_1.155}
\end{equation}
To be more precise, it is so in any $H^{+}$-algebra. In this way,
the elements of group algebra become operators. By convention, we
can choose the left regular translations. Group algebra becomes represented
by algebra of linear operators. However, this representation is badly
reducible. Namely, it is not only so that any ideal, in particular,
any minimal ideal $M(\alpha)$, is invariant under left (and right
too) regular translations (\ref{eq_1.155}). But within any minimal
two-side ideal $M(\alpha)$ spanned by all $\varepsilon(\alpha)_{mn}$
functions, separately any minimal left ideal $M(\alpha,n)$ is also
closed under all translations, and this representation is irreducible.
The minimal left ideal $M(\alpha,n)$ is spanned by all functions
$\varepsilon(\alpha)_{mn}$, where $n$ is fixed. Symmetrically, any
minimal right ideal $M(n,\alpha)$, spanned by all functions $\varepsilon(\alpha)_{nm}$
with a fixed $n$, is a representation space of some irreducible
representation of $H^{+}$ algebra. Roughly speaking, $M(\alpha,n)$, $M(n,\alpha)$ are respectively columns and rows of the matrices
$\left[\varepsilon(\alpha)_{ab}\right]$. And the elements of $M(\alpha,n)$
are "wave functions", "state vectors" of the " $\alpha$-th
type". Subspaces $M(\alpha,n_{1})$, $M(\alpha,n_{2})$,
$n_{1}\neq n_{2}$, are merely different representatives, just equivalent
descriptions of physically the same situation. By convention we may
simply fix $n=1$. And then the linear shell of all subspaces $M(\alpha,1)$
is the space of "state vectors". 

\item Another scheme is one in which $G$ is meant as the classical configuration
space (take, e.g., ${\rm SO}(3,\mathbb{R})$ or ${\rm SU}(2)$ as the configuration space of rigid body). Then all functions on $G$ are interpreted as
wave functions, and $\varepsilon(\alpha)_{mk}$, $\varepsilon(\alpha)_{ml}$
for $k\neq l$ are different wave functions, different physical situations;
one can easily construct Hamiltonians which predict "quantum
transitions" between those subspaces. But also
within such Schr\"{o}dinger wave-mechanical framework, group-algebraic
structures are physically relevant. Namely, the most important, geometrically
distinguished operators are unitaries (\ref{eq_1.93a}), (\ref{eq_1.93b}) representing transformation groups motivated by $G$. They describe some physically significant
unitary representations of $G$. And then the linear shells (\ref{eq_1.100}),
(\ref{eq_1.104}), (\ref{eq_1.111}) and so on of these representations appear
in a natural way. Roughly speaking, those linear shells are physically
interpretable representations of the abstract group algebra over $G$.
And they in a sense "parametrize" the algebra of operators acting on wave functions. If the function
$F$ in (\ref{eq_1.100}), (\ref{eq_1.104}), (\ref{eq_1.111}) is of the
$L^{1}(G)$-class or $L^{1}\left(G\times G\right)$-class, then the resulting
operators are bounded. But it is just certain unbounded operators
that are physically interesting. They describe important physical quantities.
We obtain them from the group-algebraic scheme, formally admitting
in (\ref{eq_1.100}), (\ref{eq_1.104}), (\ref{eq_1.111}) distributions
instead of functions $F$. In differential theory, when $G$ is a Lie
group, some derivatives of Dirac delta are then used. But even some
important bounded operators, e.g., identity operator and $G$-translations,
are expressed in terms of distributions, namely, Dirac deltas, as formally
included into group algebra. Differential concepts are not used then.

Physically relevant and operationally interpretable quantities are
to be sought first of all among elements of the group algebras (\ref{eq_1.100}),
(\ref{eq_1.104}), (\ref{eq_1.111}), formally extended by admitting
distributions.
\end{enumerate}

\section*{Acknowledgements}

This paper partially contains results obtained within the framework of the research project 501 018 32/1992 financed from the Scientific Research Support Fund in 2007-2010. The authors are greatly indebted to the Ministry of Science and Higher Education for this financial support. The support within the framework of Institute internal programme 203 is also greatly acknowledged.

\end{document}